\documentclass[prl,twocolumn]{revtex4}%
\usepackage{amsmath}
\usepackage{amsfonts}
\usepackage{amssymb}
\usepackage{graphicx}%
\begin{document}

\title{Novel electrically resonant terahertz metamaterials }
\author{W. J. Padilla}
\email{willie@lanl.gov} \affiliation{Los Alamos National
Laboratory, MS G756, MST-CINT, Los Alamos, NM 87545.}

\author{M. T. Aronsson}
\affiliation{Los Alamos National Laboratory, MS H851, ISR-6, Los
Alamos, NM 87545.}

\author{C. Highstrete}\altaffiliation{Also at: Dept. of Physics and Astronomy, University of New Mexico
800 Yale Blvd NE Albuquerque, New Mexico USA 87131.} \author{Mark
Lee} \affiliation{Sandia National Laboratories, P.O. Box 5800,
Albuquerque, New Mexico 87185-1415, USA}

\author{A. J. Taylor} \author{R. D. Averitt}
\affiliation{ Los Alamos National Laboratory, MS G756, MST-CINT, Los
Alamos, NM 87545.}

\begin{abstract}
We present a new class of artificial materials which exhibit a
tailored response to the \textit{electrical} component of
electromagnetic radiation. These electric metamaterials (EM-MMs) are
investigated theoretically, computationally, and experimentally
using terahertz time-domain spectroscopy. These structures display a
resonant response including regions of negative permittivity
$\epsilon_1(\omega) < 0$ ranging from $\sim$500 GHz to 1 THz.
Conventional electric media such as distributed wires are difficult
to incorporate into metamaterials. In contrast, these new localized
structures will simplify the construction of future metamaterials -
including those with negative index of refraction - and will enhance
the design and fabrication of functional THz devices.
\end{abstract}

\maketitle

Shaped dielectric and conducting materials which control the
electric component of electromagnetic fields with a designed
response have been known for many
decades.\cite{kock,bracewell,rotman} Recently, these ``artificial
dielectrics" have found renewed interest in the burgeoning field of
electromagnetic metamaterials (EM-MMs).\cite{smith1} Excitement in
EM-MMs stems from the ability of these materials to exhibit an
electromagnetic response not readily available in naturally
occurring materials including, as examples, negative refractive
index,\cite{veselago,shelby} and artificial magnetism\cite{pendry1}.
However, such exotic phenomena only became possible following the
realization that artificial materials could be designed to exhibit
an effective material response to electric\cite{pendry2} and
magnetic fields\cite{pendry1}. To date, artificial magnetic
metamaterials have been experimentally demonstrated over several
decades of frequency ranging from radio frequencies\cite{wiltshire}
to THz\cite{padilla} and near infrared frequencies\cite{costas}.

The most common element utilized for magnetic response is the
split ring resonator (SRR). Since SRRs were first used to create
negative index media, important advances have been realized.
Researchers have demonstrated designs with higher
symmetry\cite{pendryIR}, non-planar structures\cite{starr}, and
generalization to two\cite{shelby} and three
dimensions\cite{olivier}. In contrast, purely electric
metamaterials have experienced little improvement over the past 60
years and conducting wires have primarily been the medium of
choice. Wires have a potential for some tunability, i.e. a
modified plasma frequency can be obtained by making extremely thin
wires\cite{pendry2} or by adding loops thus increasing their
inductance.\cite{smith-apl} However, recent research has shown
that wire arrays are not desirable in many
ways,\cite{schurig,elec_comment}. Limiting factors include the
necessity of inter unit cell connections and specific surface
terminations\cite{fernandez}. For many applications, it would be
preferable to have a localized particle with finite extent from
which one could construct materials with an electric response.

In principle, the SRR can also be used as an electrically resonant
particle as it exhibits a strong resonant permittivity at the same
frequency as the magnetic resonance.\cite{padilla2} However, the
electric and magnetic resonant responses are coupled resulting in
rather complicated bianisotropic electromagnetic
behavior.\cite{marques,padilla3} The development of more symmetric
designs, which can be predicted by group theoretical methods,
eliminate any magneto-optical coupling effects related to
bianisotropy and yield electrically resonant
structures.\cite{padilla1} Furthermore, in the symmetric particles
the magnetic response is suppressed. Thus, such elements would
function as localized particles from which one can construct a
purely electrical resonant response.

In this letter we describe a series of new uniaxial and biaxial
electric metamaterials. The design of these symmetric structures
accomplishes the goal of creating a class of sub-wavelength
particles which exhibit a resonant response to the electric field
while minimizing or eliminating any response to the magnetic field.
Planar arrays of these new structures targeted for the THz frequency
regime have been simulated, fabricated and characterized in
transmission. Each of the EM-MMs structures show a resonant response
including a region negative permittivity $\epsilon_1(\omega) < 0$.
We discuss the advantages of these localized particles in comparison
to conventional wire-segment electric media. In particular, these
new structures will ease the burden of fabricating new metamaterials
devices including those exhibiting a negative index of refraction.

Our EM-MMs are fabricated in a planar array by conventional
photolithographic methods and consist of 200 nm thick gold with a 10
nm thick adhesion layer of titanium on semi-insulating gallium
arsenide (GaAs) substrates of 670 $\mu$m thickness. All of the
EM-MMs have an outer dimension of 36 $\mu$m, a lattice parameter of
50 $\mu$m, a line width of 4 $\mu$m, and a gap of 2 $\mu$m. The
arrays are characterized utilizing THz time domain spectroscopy
(THz-TDS) at normal incidence. The time dependence of the electric
field is measured and the complex transmissivity is obtained from
which we calculate the complex dielectric function
$\widetilde{\epsilon}(\omega)=\epsilon_1+i\epsilon_2$.\cite{padilla2}
The group-theoretical analysis \cite{padilla1} is supplemented with
finite element numerical simulations of the EM-MMs using commercial
code and Lorentz oscillator fits to the data.

\begin{figure} [ptb]
\begin{center}
\includegraphics[
width=3.25in,keepaspectratio=true
]%
{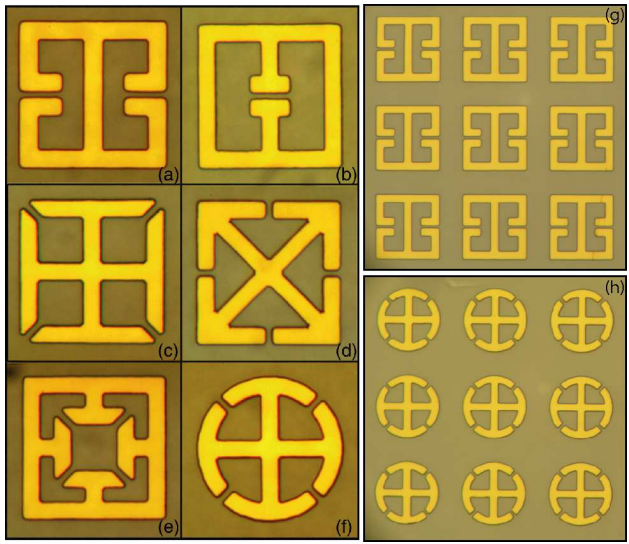}%
\caption{Photographs of the electric metamaterials characterized in
this study. The pictures of individual unit cells of EM-MMs in (a)
through (f) are termed, in the text, E$_{1}$-E$_{6}$ respectively.
In (a) and (b) we show EM-MMs with a resonant uniaxial electric
response when the electric field is polarized vertically. (c)
through (f) show EM-MMs which exhibit a biaxial electric response
where the electric field can be polarized vertically, horizontally,
or unpolarized. In (g) and (h) photographs with an expanded view of
the samples shown in (a) and (f) demonstrate how the individual
particles are arranged into an array.}%
\label{fig1}%
\end{center}
\end{figure}

\begin{figure*}
[ptb]
\begin{center}
\includegraphics[
width=6.0in,keepaspectratio=true
]%
{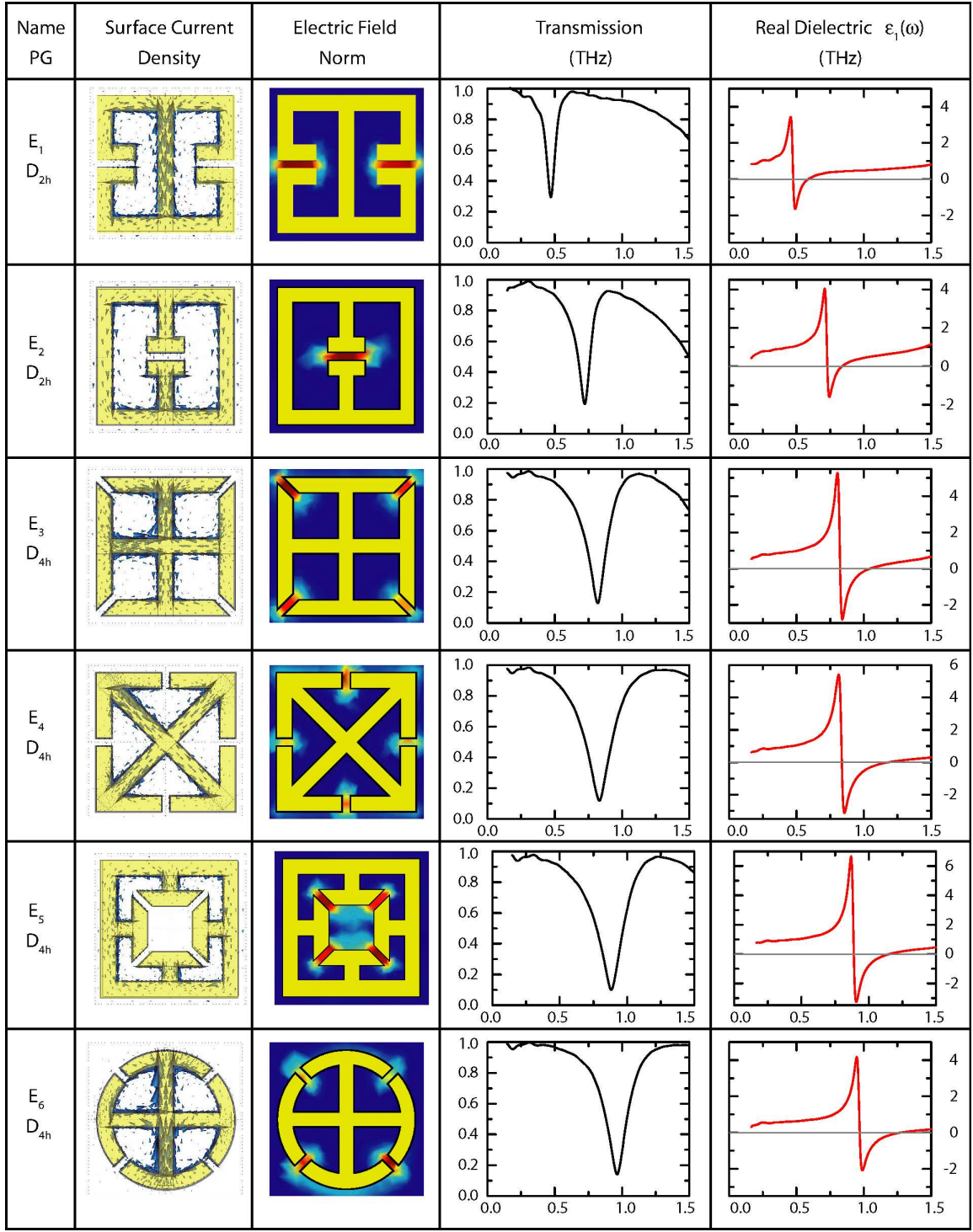}%
\caption{Simulation and experimental results for new electric
metamaterial particles. The left column lists the names of particles
as we address them in this article and the point group in Shoenflies
notation. The second and third columns show the surface current
density and norm of the electric field at resonance, respectively.
The last columns show experimental the experimentally measured
transmission T($\omega$) and real part of the dielectric function
$\epsilon(\omega)$.}%
\label{fig2}%
\end{center}
\end{figure*}

In Fig. \ref{fig1} we show photographs of the electric metamaterials
characterized in this study. As mentioned above, each of these
particles is designed \cite{padilla1} to exhibit a resonant response
to the electric field while minimizing or eliminating any response
to the magnetic field. In Fig. \ref{fig2} we show simulation and
experimental results. The second column shows the calculated surface
current density which provides a simple way to visualize the absence
of a magnetic response. Namely, the magnetic fields created by
circulating surface currents cancel due to clockwise and
counter-clockwise components in adjacent regions of the particle.
Thus, any resonant response must necessarily be of electrical origin
since there is no net circulation of current in each unit cell. The
third column shows the norm of the electric field at resonance. The
red regions in the gap indicate a strong local field enhancement
which, according to the simulations, can be upwards of $10^{4}$ of
the incident field. The last two columns of Fig. \ref{fig2} show the
experimentally measured field transmission T($\omega$) and real part
of the dielectric function $\epsilon_1(\omega)$, respectively. Each
structure exhibits a very strong resonance with the transmission
decreasing to as little as $\sim$10 percent. Additionally, all of
the EM-MMs characterized in this study display regions of negative
permittivity.

In Table I we summarize some characteristic parameters related to
the $\epsilon(\omega)$ response. A Drude-Lorentz
model\cite{wooten} was used to fit the $\epsilon_1(\omega)$ data,
from which we extract $\omega_0$ which is the center frequency,
and $\omega_p$ which is the frequency of the zero-crossing of
$\epsilon_1(\omega)$. In addition, we list the minimum value of
$\epsilon_{1}$, the oscillator strength
($S=\omega_p^2/\omega_0^2$), the percentage bandwidth over which
$\epsilon_1 < 0$ is achieved, and the ratio of the free space
wavelength to the unit cell length $\lambda_0$/a.

\begin{table}[hbtp]\label{table1}
\centering
\begin{tabular}{|c|c|c|c|c|c|c|}
\hline
Name & $\omega_0$(THz) & $\omega_p$(THz) & min $\epsilon_1$ & S=$\omega_p^2/\omega_0^2$ & BW(\%) & $\lambda_0$/a\\
\hline
E$_1 $    & 0.480 &  0.587  & -1.63 & 1.50 & 22.3 & 12.5\\
\hline
E$_2 $    & 0.730 &  0.837  & -1.59 & 1.31 & 14.7 & 8.2\\
\hline
E$_3 $    & 0.826 &  1.046  & -2.80 & 1.60 & 26.6 & 7.3\\
\hline
E$_4 $    & 0.840 &  1.178  & -3.12 & 1.97 & 40.2 & 7.1\\
\hline
E$_5 $    & 0.892 &  1.152  & -3.26 & 1.67 & 29.1 & 6.7\\
\hline
E$_6 $    & 0.972 &  1.270  & -2.08 & 1.71 & 30.7 & 6.2\\
\hline
\end{tabular}
\caption{Key parameters quantifying the electric response of
metamaterials characterized in this study. Columns 2 and 3 show the
center frequency of the oscillator $\omega_0$ and the zero crossing
of the epsilon response $\omega_p$, respectively. The fourth column
lists the minimum value of $\epsilon_1$ and the fifth column the
oscillator strength. Column six displays the region over which
$\epsilon_1 < 0$ is achieved in terms of band width (BW) normalized
by $\omega_0$. The last column lists the ratio of the free space
wavelength at resonance ($\lambda_0\sim\omega_0^{-1}$) to the unit
cell length (a=50 $\mu$m), an indication of how deep into the
effective medium regime the materials are.}
\end{table}

Next we compare these new electrically resonant particles with wires
which are the canonical electric metamaterial
\cite{pendry2,smith-apl}. As mentioned, typical electric
metamaterials utilize straight wires which can be thought of as a
bulk metal with a reduction in the effective electron density due to
the reduction in volume fraction of the metal. A bulk metal has free
electrons that screen external EM fields from penetrating inside the
material for frequencies below the plasma frequency defined as
$\omega_p^2=4\pi n e^2/m^*$, where $n$ is the effective electron
density and $m^*$ is the effective mass. For an electric field
polarized along the wire axis an effective medium response is
obtained with a modified plasma frequency. Wires can also be formed
with loops to add additional inductance to increase carrier mass and
reduce the plasma frequency.\cite{smith-apl} Lastly, cuts can be
added periodically along the wires to obtain a Drude-Lorentz
response for added tunability. However, as discussed above, wires
have disadvantages which can limit their functionality as electric
metmaterials.

We now highlight several advantageous features of these new electric
metamaterials which principally derive from their symmetry and
localized extent. For THz metamaterials with characteristic lengths
a=50$\mu$m, the samples are easily fabricated with standard optical
lithographic methods and a simple normal incidence transmission
measurement is all that is required to characterize their full
electromagnetic behavior. Issues related to connectivity (as for
wires) do not arise since the resonant response derives from
particles within individual unit cells. Wires in two and three
dimensions, from a fabrication viewpoint, are extremely difficult to
implement. However, the electric structures presented here
generalize to higher dimensions in a simple and straightforward
manner, similar to SRRs.\cite{pendry1} As mentioned above, E$_{1}$ -
E$_{6}$ do not exhibit a magnetic response near the resonant
frequency (i.e. $\mu=1$). Thus, these particles are natural
complements to magnetically active SRRs meaning that it is possible
to create negative index materials through appropriate combinations
of these two varieties of sub-wavelength particles.

Although the thickness of the samples characterized in this study
was relatively thin (200 nm), E$_{1}$ - E$_{6}$ yielded regions of
negative permittivity and decent bandwidth. We note that there are
several simple ways to improve the response of the EM-MMs in this
study, namely: thicker samples, increased filling fraction,
extension to multiple layer structures, and utilization of higher
conductivity metals. However, the results presented in Figure
\ref{fig2} and Table I clearly show that the present metamaterials
already display a pronounced and functional terahertz response
which, when combined with magnetically resonant SRRs, will
facilitate a new approach to creating negative index metamaterials.

At terahertz frequencies there is a lack of intrinsic response
from natural materials, known as the ``THz gap".\cite{gwyn} Taking
advantage of this void in THz electromagnetic material response is
desirable for many potential applications such as: personnel and
luggage screening, explosives detection, and all weather imaging.
The initial demonstration of several new electric metamaterials at
THz frequencies highlights their usefulness and versatility. These
new structures can be expected to play an important role filling
in the THz gap.

In conclusion, we have presented new designs for metamaterials
that exhibit a tailored resonant electrical response investigated
using THz time domain spectroscopy. The samples offer significant
advantages over current electric metamaterials, both in terms of
fabrication as well as characterization. Each of the EM-MMs
characterized in this study exhibit a negative dielectric
response, which may be useful for future devices. Further, EM-MMs
may be constructed to exhibit a polarization sensitive response.
These electric metamaterials will significantly ease the burden of
construction for future negative index metamaterial devices, and
their initial demonstration at THz frequencies highlights their
potential as functional electromagnetic materials.

W. J. P. acknowledges support from the Los Alamos National
Laboratory LDRD program. We also acknowledge support from the Center
for Integrated Nanotechnologies. Sandia is a multiprogram laboratory
operated by Sandia Corporation, a Lockheed Martin Company, for the
United States Department of Energy's National Nuclear Security
Administration under Contract No. DE-AC04-94AL8500.

\end{document}